\documentclass[12pt]{article}
\usepackage[a4paper,margin=1in]{geometry}
\usepackage{graphicx,verbatim} 
\usepackage{amsthm}

\usepackage{amssymb}
\usepackage{booktabs}

\usepackage{amsmath}
\usepackage{hyperref}
\usepackage{mdframed,enumerate}
\usepackage{tabularx}

\title{Agentic AI for Cyber Resilience: A New Security Paradigm and Its System-Theoretic Foundations}
\author{Tao Li and Quanyan Zhu\thanks{T. Li is with the Department of Systems Engineering, City University of Hong Kong; E-mail: 	li.tao@cityu.edu.hk; Q. Zhu is with the Department of Electrical and Computer Engineering, New York University; E-mail: qz494@nyu.edu. Correspondence should be addressed to T. Li. }}

\date{}

\begin{document}

\maketitle

\begin{abstract}
Cybersecurity is being fundamentally reshaped by foundation-model--based
artificial intelligence.
Large language models now enable autonomous planning, tool orchestration, and
strategic adaptation at scale, challenging security architectures built on static
rules, perimeter defenses, and human-centered workflows. This chapter argues for a shift from prevention-centric security toward
\emph{agentic cyber resilience}.
Rather than seeking perfect protection, resilient systems must anticipate
disruption, maintain critical functions under attack, recover efficiently, and
learn continuously.
We situate this shift within the historical evolution of cybersecurity paradigms,
culminating in an \emph{AI-augmented paradigm} where autonomous agents participate
directly in sensing, reasoning, action, and adaptation across cyber and
cyber--physical systems. We then develop a system-level framework for designing agentic AI workflows.
A general agentic architecture is introduced, and attacker and defender workflows are
analyzed as coupled adaptive processes, and game-theoretic formulations are shown
to provide a unifying design language for autonomy allocation, information flow,
and temporal composition.
Case studies in automated penetration testing, remediation, and cyber deception
illustrate how equilibrium-based design enables system-level resiliency design. 
\end{abstract}

\section{Introduction}

Cybersecurity is currently undergoing  a new wave of disruptive changes.
The rapid emergence of \emph{foundation-model–based AI}, including large language
models and multimodal reasoning systems, has altered not only the scale of cyber
threats but their fundamental character.
These models enable reasoning, planning, and tool orchestration at a level of
generality and flexibility that was previously unavailable.
Attackers can now automate reconnaissance, generate adaptive exploits, and
coordinate campaigns using foundation models as cognitive engines, while
defenders face increasing system complexity, compressed decision timelines, and
the limits of static rule-driven security architectures.
This shift exposes structural mismatches between traditional security models and
an environment shaped by general-purpose agentic intelligence.

Responding to such a disruption requires more than incremental improvements to
existing tools.
It calls for reexamining the foundational assumptions that govern how security is
designed, operated, and integrated into complex socio-technical systems \cite{zhu2025revisiting,huang2023cognitive}.
By revisiting the historical evolution of cybersecurity paradigms, we can clarify
why prior approaches struggle under foundation-model–driven threats and what
conceptual shifts are required to accommodate autonomous reasoning, strategic
adaptation, and continuous interaction between attackers and defenders.
At the same time, this historical perspective provides a basis for envisioning
future security architectures that are resilient by design rather than brittle
under change.

The following subsection therefore traces the evolution of cybersecurity through
a sequence of paradigms, from early laissez-faire protection to the emerging
AI-augmented, agentic, and resilience-oriented model.
This perspective situates today's disruption, driven by the foundationmodel, within a
broader conceptual trajectory and motivates the shift toward agentic cyber
resilience that frames the remainder of this chapter.

\subsection{From Laissez-faire to AI-Augmented Paradigms}
The evolution of cybersecurity can be understood as a sequence of distinct
\emph{paradigms}, progressing from early ad hoc protection to the emerging
\emph{AI-augmented paradigm} \cite{huang2022ai}.
Each paradigm reflects a fundamentally different worldview regarding how security
is conceptualized, implemented, and governed within socio-technical systems.
Transitions between paradigms have been driven by the co-evolution of computing
infrastructure, threat sophistication, and the shifting allocation of
decision-making authority between humans and machines.

Viewed through a historical lens, these paradigms follow a clear temporal
trajectory:
\begin{itemize}
\item \textbf{1P (1970s--1990s):} Laissez-faire security for isolated systems
\item \textbf{2P (1990s--early 2000s):} Perimeter-based defense for networked systems
\item \textbf{3P (2000s--2010s):} Reactive detection and incident response
\item \textbf{4P (2010s--early 2020s):} Proactive, intelligence-driven defense
\item \textbf{5P (2020s--):} AI-augmented, agentic, and resilience-oriented security
\end{itemize}
Although these paradigms frequently coexist in practice, each represents a
qualitative shift in how security is understood, operationalized, and valued.

The \textbf{first paradigm}, the \emph{laissez-faire paradigm (1P)}, was characterized
by minimal regulation and largely reactive postures.
Computing systems were relatively isolated, threats were assumed to be rare, and
security was treated as a secondary engineering concern.
Incident response was manual, localized, and slow, and resilience was not an
explicit design consideration.

The \textbf{second paradigm}, the \emph{perimeter paradigm (2P)}, emerged as networks
expanded and organizational boundaries became digitally porous.
Security architectures focused on separating trusted internal systems from
untrusted external networks through firewalls, access controls, and intrusion
prevention mechanisms.
While this approach improved scalability, it proved structurally fragile: once the
perimeter was breached, internal systems were often insufficiently protected.

The \textbf{third paradigm}, the \emph{reactive paradigm (3P)}, arose in response to
the increasing frequency and sophistication of cyber attacks.
Defenders shifted from static protection to continuous monitoring and detection,
deploying intrusion detection systems, antivirus software, and SIEM platforms.
Security operations centers became institutionalized as hubs for incident response.
However, this paradigm remained fundamentally reactive and human-centric,
characterized by alert fatigue, delayed response, and a tendency to treat attacks
as isolated events rather than sustained adversarial campaigns.

The \textbf{fourth paradigm}, the \emph{proactive paradigm (4P)}, marked a transition
from reaction toward anticipation.
Threat intelligence, risk modeling, red–blue team exercises, and threat hunting
were introduced to infer attacker intent and disrupt campaigns earlier in their
lifecycle.
Despite these advances, proactive defense continued to rely heavily on expert
judgment, hand-crafted heuristics, and static models, limiting scalability and
adaptivity in the presence of fast-moving and AI-enabled adversaries \cite{pawlick2017proactive,pawlick2017strategic,ge2023zero,ge2023scenario,jajodia2016cyber}.

The \textbf{fifth paradigm}, the \emph{AI-augmented paradigm (5P)}, constitutes a
fundamental departure from prior approaches.
Its defining innovation is not merely the use of AI for automation, but the shift
from human-supervised workflows to \emph{AI-orchestrated, agentic security
ecosystems}.
Artificial intelligence, machine learning, and large language models operate across
the full security lifecycle—from sensing and detection to response, recovery, and
post-incident learning \cite{li2025texts,li2024symbiotic}.
Security is no longer treated as a static configuration problem, but as a
continuous \emph{sense--reason--act} process executed by autonomous,
learning-enabled systems.

\subsection{AI-Augmented Security Paradigm}
What distinguishes 5P most clearly is the emergence of \emph{agentic autonomy and
adaptivity}.
Security functions evolve from rule-based execution toward goal-directed behavior,
where learning agents dynamically adjust strategies in response to uncertainty,
feedback, and adversarial adaptation.
Decisions arise from closed-loop optimization rather than fixed policies.
At the same time, AI enables deep integration across traditionally siloed domains,
bridging detection, forensics, patching, access control, and cyber–physical
operations within a shared cognitive architecture.

Beyond operational efficiency, the fifth paradigm introduces cognitive and
strategic awareness into cybersecurity.
AI-augmented systems reason about intent, simulate adversarial behavior, and
anticipate escalation, deception, and long-term adaptation.
Defense becomes predictive and game-aware, shaped by ongoing interaction among
intelligent agents on both sides.
Rather than a monolithic defender, the security ecosystem consists of distributed,
communicating, and self-organizing \emph{agentic entities} whose collective behavior
produces defensive intelligence.

Equally important, 5P reorients cybersecurity toward resilience \cite{zhu2025foundations,yang2024game}.
Instead of striving for perfect prevention, systems are designed to anticipate
disruption, absorb shocks, recover functionality, and learn from failure.
Resilience becomes a first-class design objective, reflecting the reality that
absolute security is neither achievable nor economically sustainable.
Through embedded learning and self-healing mechanisms, AI-augmented systems sustain
mission continuity under persistent and adaptive threats.

In this sense, the fifth paradigm integrates security and resilience into a single
adaptive process \cite{zhu_basar_resilience2024}.
Where earlier paradigms sought to minimize risk through static controls, 5P
embraces uncertainty and operates within a co-evolutionary landscape of intelligent
adversaries.
Security becomes a dynamic game of reasoning and adaptation among autonomous
agents.
The AI-augmented paradigm thus serves as the conceptual bridge to
\emph{agentic cyber resilience}, where artificial intelligence functions not merely
as a support tool, but as a strategic actor embedded in the ongoing defense,
recovery, and evolution of complex socio-technical systems.

\subsection{Structure of the Chapter}

This chapter builds on the vision of the AI-augmented paradigm by developing a
\emph{system-level theory for agentic cyber resilience}.
We begin in Section 2 by introducing a general architectural model for agentic AI workflows,
clarifying the roles of memory, reasoning, tool invocation, human interaction, and
environmental embodiment.
We then examine how these architectural elements are instantiated within cybersecurity
settings, with particular attention to advanced persistent threats and
adversarial learning dynamics. In Chapter 3, we present the cyber kill chain under agentic AI, showing how the attacker and
defender workflows become tightly coupled through continuous feedback and
adaptation.
We conceptualize this interaction as a strategic game among learning agents, drawing
on concepts from game theory and control to reason about stability, incentives,
and long-term behavior. The chapter then shifts from security to resilience in Section 4, discussing the temporal
facets of proactive, responsive, and retrospective resilience, and explaining in Section 5
the frameworks for the design of agentic AI workflows.
Finally, in Section 6, we discuss integrated cyber–physical systems, where
agentic AI and physical AI jointly govern both digital and embodied processes.

\section{Agentic AI for Security: Toward Agentic Defense Workflows}
\label{sec:agentic_security}

\begin{figure}[t]
    \centering
    \includegraphics[width=\linewidth]{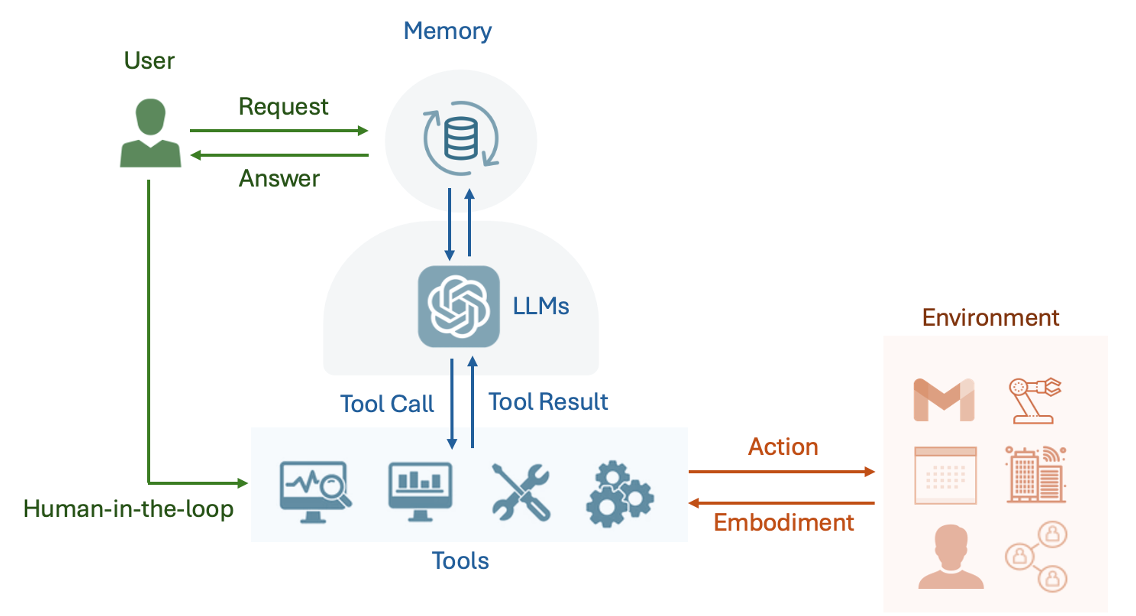}
    \caption{\textbf{A general agentic AI architecture.}
    The figure depicts a canonical agentic system composed of a reasoning
    core (LLMs), persistent memory, tool interfaces, human-in-the-loop
    interaction, and an external environment.
    User requests are processed through memory-aware reasoning; the agent
    invokes tools to perceive and act upon the environment, receives
    structured feedback, and updates its internal state through learning.
    This closed-loop architecture supports long-horizon reasoning,
    adaptive behavior, and embodied interaction across digital and
    physical domains.}
    \label{fig:agentic_general}
\end{figure}

AI now plays a transformative role in complex decision-making systems by moving beyond narrow prediction and classification toward integrated cognition and action.
On one hand, AI assists human operators in tasks such as monitoring, analysis, and decision support, for example, by summarizing large data streams, flagging anomalies, or recommending candidate actions.
On the other hand, recent advances in learning, reasoning, and autonomy increasingly enable AI systems to \emph{plan, decide, and act} with limited or intermittent human intervention across complex, multi-layered environments.
This evolution marks a fundamental transition from tool-centric automation, where AI executes predefined functions, to \emph{agentic AI}, in which AI systems are designed as goal-directed entities that continuously interact with humans, tools, and the external world.

Figure~\ref{fig:agentic_general} presents a \emph{general architectural template for agentic AI}, independent of specific application domains.
The diagram organizes agentic intelligence into a small number of interacting blocks, each of which plays a distinct functional role.

\paragraph{Reasoning Core (LLMs).}
At the center of the architecture is a reasoning engine, instantiated here by
large language models (LLMs).
This component performs high-level interpretation, planning, and synthesis: it
translates user requests and observations into internal representations,
generates candidate plans, and reasons about multi-step courses of action.
Unlike traditional pipelines that map inputs directly to outputs, the reasoning
core supports deliberation, abstraction, and goal-conditioned behavior.
Recent work on reasoning- and tool-augmented LLMs demonstrates that such models
can decompose complex tasks, reason over intermediate states, and adapt plans
based on feedback rather than executing static scripts
(e.g., ReAct, Toolformer, and subsequent agent frameworks)~\cite{yao2022react,schick2023toolformer}.

\paragraph{Persistent Memory.}
Above the reasoning core, the memory block provides a persistent state across
interactions.
This memory stores historical exchanges with users, prior tool results,
contextual knowledge about the environment, and learned abstractions about the
agent’s own behavior.
Persistent memory enables temporal continuity: the agent can recognize recurring
situations, accumulate experience, and reason over long horizons rather than
reacting myopically to the current input.
In practice, this memory may be implemented using vector databases for episodic
retrieval, symbolic knowledge graphs for structured reasoning, or hybrid
episodic–semantic memory systems~\cite{lewis2020retrieval,kumar2024supporting}.
Such mechanisms are essential for situational awareness, personalization, and
continual learning from past successes and failures.

\paragraph{Tool Interfaces.}
The lower block of the diagram represents the agent’s interface to external tools.
Rather than directly manipulating the environment, the agent issues structured
\emph{tool calls} to sensing, analytics, simulation, optimization, or control
modules, and receives structured \emph{tool results} in return.
Examples include querying monitoring dashboards, invoking data analysis
pipelines, running simulations or digital twins, or triggering automated control
actions.
This paradigm, exemplified by modern agent frameworks, grounds the agent’s
reasoning in real system state and measurable outcomes rather than purely
symbolic inference~\cite{schick2023toolformer,wu2024autogen}.
The feedback loop formed by tool invocation, result interpretation, and memory
update constitutes the computational substrate for adaptive, goal-directed
behavior.

\paragraph{Human-in-the-Loop.}
A defining feature of the architecture is the explicit inclusion of
human-in-the-loop interaction.
Humans may issue high-level requests, refine objectives, provide corrective
feedback, supervise tool execution, or intervene in critical decisions.
Rather than remaining external supervisors, human operators are integrated as
first-class components of the agentic loop.
This integration supports alignment, accountability, and trust, particularly in
high-stakes settings such as cybersecurity, healthcare, and critical
infrastructure.
Importantly, the architecture supports adjustable autonomy: agents may operate
independently under routine conditions, while deferring to human judgment when
uncertainty, risk, or ethical considerations exceed predefined thresholds, a
principle emphasized in human–AI teaming and mixed-initiative control
literature~\cite{amershi2019guidelines}.

\paragraph{Environment and Embodiment.}
On the right side of the diagram, the environment represents the external world
with which the agent interacts.
This environment may include digital systems (e.g., software services,
networks), physical systems (e.g., robots, industrial processes), organizational
processes, and other agents.
Actions taken by the agent, mediated through tools, alter the environment, while
observations of those changes are fed back via sensing and analytics.
This embodiment closes a continuous sense–reason–act–learn loop, distinguishing
agentic AI from static automation or batch learning systems.
By embedding cognition within ongoing interaction with the world, agentic AI
supports adaptation under uncertainty and nonstationarity, aligning with
foundational perspectives in embodied intelligence and adaptive control
systems~\cite{brooks1991intelligence}.

\paragraph{Integration.}
The components in Fig.~\ref{fig:agentic_general} form a closed-loop architecture
in which reasoning, memory, tools, human oversight, and environmental interaction
are tightly coupled.
The reasoning core generates plans and actions; tools ground those plans in
operational reality; memory provides continuity and learning; humans ensure
alignment and accountability; and the environment supplies feedback that drives
adaptation.
This architecture is intentionally domain-agnostic, serving as a general
template for agentic systems across application areas, and providing the
conceptual foundation for instantiating agentic AI in adversarial and
resilience-critical domains such as cybersecurity.

Although Fig.~\ref{fig:agentic_general} is   domain-agnostic, it provides a powerful conceptual foundation for applying agentic AI to cybersecurity and defense.
When instantiated in security settings, each architectural block acquires a concrete operational meaning grounded in adversarial environments, where intelligent attackers probe, adapt, and persist over time. In this context, the \emph{environment} corresponds to enterprise networks, cloud infrastructures, endpoints, users, and cyber–physical systems such as industrial control systems or autonomous platforms \cite{xu2015secure,jin2025cloud,xu2018cross,zhu2011hierarchical,rieger2019industrial,zhu2021cybersecurity}.
The state of this environment evolves continuously as legitimate users interact with systems, configurations change, and adversaries conduct reconnaissance, exploitation, and lateral movement.
For example, in an advanced persistent threat (APT) intrusion \cite{zhu2018multi,zhu2011robust,huang2019dynamic,huang2020dynamic,li24col-aisg, hammer25col}, early-stage reconnaissance may manifest as subtle shifts in network traffic distributions, anomalous authentication attempts against privileged accounts, or unusual process executions that individually appear benign but collectively signal adversarial intent.
An agentic defense system treats these changes as part of a dynamic environment rather than isolated events, continuously updating its situational awareness as the intrusion unfolds.

The \emph{tool interfaces} in Fig.~\ref{fig:agentic_general} map naturally to the
diverse ecosystem of cybersecurity instrumentation and control mechanisms.
These include network- and host-based monitors, intrusion detection and prevention
systems, log aggregation and analytics platforms, vulnerability scanners,
sandboxing environments, and response mechanisms such as access-control
reconfiguration or service isolation~\cite{singer2013cybersecurity,scarfone2007guide}.
Through structured tool calls, an agentic AI can query measurements, invoke
simulations to assess the impact of potential actions, or execute containment
measures.
For instance, during an advanced persistent threat (APT) campaign, the agent may
call a network flow analyzer to determine whether low-volume outbound traffic
corresponds to covert data exfiltration, or invoke a sandbox to analyze a
suspicious binary before deciding whether to isolate the affected host.
Such tool-mediated interactions ground the agent’s reasoning in measurable system
states and observable outcomes, a critical requirement for operating in
adversarial environments~\cite{yao2022react,schick2023toolformer}.

The \emph{persistent memory} component plays a critical role in adversarial
reasoning.
In cybersecurity, memory encodes attack histories, indicators of compromise,
system configurations, and learned threat models accumulated across incidents.
By retaining contextual knowledge over time, the agent can recognize recurring
attack patterns, correlate seemingly isolated alerts into coherent campaigns, and
learn from past encounters~\cite{mitre_attack}.
In the APT setting, memory enables the agent to associate current activity, such as
credential reuse or privilege escalation, with earlier reconnaissance phases
observed weeks or months prior, thereby identifying persistence strategies that
would evade short-term detection mechanisms.
This form of temporal continuity is essential for defending against adversaries
whose defining characteristics are patience, stealth, and iterative
adaptation~\cite{alshamrani2019survey}.

At the center of the architecture, the \emph{reasoning core} synthesizes
observations from tools and memory to form hypotheses about adversarial intent and
to generate candidate defense strategies.
Rather than merely classifying events as benign or malicious, an agentic AI
reasons about sequences of actions, anticipates attacker responses, and evaluates
tradeoffs among competing objectives such as containment speed, operational
disruption, intelligence collection, and forensic value.
For example, when an APT foothold is suspected, the agent may deliberate between
immediately isolating a compromised host, which reduces risk but may reveal
detection, and allowing limited interaction within a controlled environment to
observe attacker behavior and infer objectives.
As new evidence arrives, the reasoning core revises its hypotheses and adapts its
strategy accordingly, reflecting principles studied in adversarial learning and
sequential decision-making under uncertainty~\cite{barreno2006can,huang2011adversarial}.

Human-in-the-loop interaction remains essential in security-critical deployments.
Analysts may provide high-level objectives, validate or override automated
actions, or supply contextual knowledge that is difficult to infer from data
alone, such as business criticality, regulatory constraints, or geopolitical
considerations.
An agentic defense system therefore operates with adjustable autonomy: it handles
routine detection and response actions independently, while escalating ambiguous
or high-impact situations, such as suspected compromise of mission-critical
assets, to human operators.
This mixed-initiative collaboration aligns with best practices in human--AI
teaming and supports accountability and trust in high-stakes cyber
operations~\cite{amershi2019guidelines, tao24ddztd}.

When these components are integrated, the general agentic architecture gives rise
to \emph{agentic defense workflows}.
In such workflows, AI systems do not merely analyze security data post hoc, but
actively participate in the cyber defense loop.
They continuously sense the environment through tools, reason about evolving
threats using memory and planning, act through response and deception mechanisms,
and learn from the outcomes of those actions.
In an APT scenario, this may involve autonomously detecting early reconnaissance,
deploying adaptive deception to mislead the adversary, selectively containing
exploitation attempts, and updating threat models to improve detection of future
campaigns.
Viewed through this lens, cyber defense becomes a closed-loop, learning-enabled,
and adversarially aware process rather than a static monitoring function.

\section{Agentic AI in the Cyber Kill Chain}
\label{sec:agentic_kill_chain}

Despite the fact that earlier waves of AI-enabled cybersecurity already incorporated elements of
reasoning, learning, and automation, these capabilities were typically
expensive and tightly coupled to specialized expertise.
Advanced reasoning engines, large-scale simulations, and expert-driven analytics
were accessible only to well-resourced organizations and were often confined to
isolated segments of the security workflow.
As a result, intelligent decision-making remained centralized, slow to adapt,
and limited in both scale and scope.

The emergence of agentic AI fundamentally alters this landscape.
By significantly lowering the cost of reasoning, planning, and adaptation, and by
reducing the dependence on scarce domain expertise, agentic AI makes sophisticated
decision-making broadly accessible.
Both attackers and defenders can now easily acquire and integrate agentic AI
services into their workflows, often as modular components that interoperate with
existing tools and infrastructure.
This democratization of autonomy means that agentic reasoning is no longer the
exclusive advantage of a single side; instead, it permeates the entire
cyber-ecosystem, and it enables rapid adoption on both offense and defense.

\subsection{Cyber Kill and Defense Chains}
The cyber kill chain describes a
\emph{linear} sequence of attacker actions, progressing from reconnaissance and
weaponization to exploitation and the achievement of objectives.
This abstraction is particularly well-suited for characterizing
\emph{advanced persistent threats (APTs)}, which unfold over extended time
horizons and adapt continuously in response to defensive measures.
APTs are distinguished by stealthy reconnaissance, staged and selective
exploitation, and iterative refinement of tactics based on observed system and
defender behavior.

In recent years, APT campaigns have become markedly more sophisticated as they
increasingly leverage artificial intelligence, large-scale automation, and
substantial organizational resources.
AI-enhanced reconnaissance, adaptive malware, and automated command-and-control
have expanded the speed, scale, and persistence of such operations.
At the level of nation-state actors, these capabilities blur the boundary between
cyber operations and strategic conflict, elevating APT activity from isolated
intrusions to sustained campaigns that resemble digital warfare in both intent
and impact.

When agentic AI systems participate, it is not merely the linear structure of the
kill chain that changes.
Rather, the attacker’s kill chain becomes tightly \emph{intertwined} with a
corresponding defensive chain, reflecting the co-evolutionary nature of modern
APT campaigns.
Both attackers and defenders deploy autonomous agents that sense, reason, and act
in continuous, coupled feedback loops.
Reconnaissance is met with anticipatory monitoring, exploitation with adaptive
containment, and long-term persistence with automated recovery and dynamic
reconfiguration.
The kill chain thus evolves from a one-directional sequence into a coupled
attacker–defender interaction, enabled and accelerated by agentic intelligence
on both sides.

Table~\ref{tab:killchain} illustrates this dual-use nature by showing how
agentic AI can serve simultaneously as an enabler of offense and as an architect
of defense across every stage of the cyber kill chain.
The same underlying capabilities, including autonomous exploration, contextual
reasoning, and adaptive learning, can be leveraged for exploit discovery and
exploit mitigation, social engineering, and cognitive defense, as well as
command-and-control (C2) and counter-C2 mechanisms.

On the offensive side, agentic AI enables large-scale reconnaissance, automated
vulnerability discovery, adaptive malware generation, and resilient C2
coordination~\cite{anderson2018learning,alshamrani2019survey,guo2025sok,datta2025agentic}.
On the defensive side, these same capabilities support anticipatory monitoring,
threat hunting, adaptive containment, deception, and automated recovery,
transforming defense from static protection into a continuous,
learning-driven process~\cite{barreno2006can,mitre_attack}.
This symmetry underscores that agentic AI does not inherently favor offense or
defense, but instead reshapes the cyber kill chain into a co-evolutionary arena in
which intelligent agents on both sides learn, adapt, and counter-adapt over time.

\subsection{Agentic AI Arms Race}
This dual-use nature of agentic AI raises a central question:
\emph{if both sides employ agentic AI, who will win, and how?}
The answer is nontrivial because cybersecurity is intrinsically asymmetric.
Attackers need to succeed only once to achieve impact, while defenders must succeed
continuously across time, assets, and attack vectors.
A single overlooked vulnerability, misconfiguration, or delayed response can
invalidate an otherwise robust defensive posture.
Defenders, moreover, must operate under constraints imposed by usability,
availability, regulatory compliance, and operational cost constraints that
attackers do not share.

Agentic AI does not eliminate this asymmetry; in many cases, it
\emph{amplifies} it.
Autonomous attack agents can probe vast attack surfaces, explore rare corner
cases, and rapidly mutate tactics at machine speed.
They can parallelize exploration, personalize attacks, and exploit distribution
shifts faster than human-centered defense workflows can respond.
Defensive agents, by contrast, must balance detection accuracy against false
positives, manage limited response resources, and preserve mission continuity.
Even as defensive models improve, the combinatorial space of possible attacks
continues to expand, preserving a persistent advantage for offense.

An example of such adversarial dynamics is found in adaptive malware detection and evasion, a problem explicitly addressed in recent adversarial AI defense efforts such as DARPA’s \emph{Guaranteed AI Robustness for Decision-Making (GARD)} program~\cite{darpa_gard}.
In this setting, defenders deploy machine-learning models trained on static and dynamic representations to detect malicious software at scale, while anticipating that attackers will actively probe and adapt to learned decision boundaries~\cite{barreno2006can,huang2011adversarial}.

Attackers respond by algorithmically generating obfuscated or polymorphic variants that preserve malicious functionality while shifting feature representations outside the detector’s training distribution, often optimizing evasion through automated search or reinforcement learning~\cite{anderson2018learning}.
Defenders counter by retraining models, incorporating more semantically grounded features, or deploying ensemble detection pipelines, only for attackers to adapt again by exploiting structural blind spots such as delayed execution, environmental triggers, or benign-behavior mimicry.

This detect–evade–retrain–counter-evade cycle is not an implementation flaw but a structural property of learning-based defense in adversarial environments~\cite{barreno2006can}.
It reflects a repeated interaction between intelligent agents with opposing objectives, underscoring that adversarial learning in cybersecurity must be understood as a \emph{dynamic strategic game}, rather than a one-shot classification problem. Cyber conflict becomes an \emph{arms race among learning
agents}.
Each side improves its models, tools, and strategies in response to the other.
Progress on one side induces counter-progress on the other, resulting in a
continuous race rather than a terminal victory.
This race unfolds across multiple time scales: tactical adaptation during an
incident, operational learning across campaigns, and strategic investment in
capabilities over months or years \cite{yang2025toward,yang2025multi,li2024symbiotic}.

\subsection{Game-Theoretic Frameworks and Cyber Resilience}
Game theory provides a natural and principled language for analyzing and designing such adversarial interactions \cite{manshaei2013game,zhu2018game,tambe2011security,pawlick2021game,kamhoua2021game,rass2018game,zhu2015game}.
In agentic cyber environments, attackers and defenders are not static entities but strategic decision makers whose actions influence one another over time.
Game-theoretic models capture this interdependence by explicitly representing objectives, information structures, and strategic choices under uncertainty.

At the \emph{tactical level}, games model real-time attacker–defender interactions during ongoing incidents.
Here, game formulations help reason about optimal response actions, adaptive containment policies, and the strategic use of deception, such as honeypots or moving-target defenses \cite{zhu2013game,boumkheld2019honeypot}.
Uncertainty about attacker intent, system state, and future actions can be formalized through incomplete-information or stochastic games, enabling defenders to anticipate adversarial reactions rather than responding myopically to observed events.

At the \emph{strategic level}, game-theoretic reasoning captures longer-term dynamics, including capability investment, information asymmetry, commitment strategies, and deterrence.
Repeated and dynamic games provide a framework for understanding how learning agents adapt across campaigns, how credibility and reputation emerge, and how persistent advantages or stalemates can arise over time \cite{zhang2021informational, tao23transparent, tao23pot,zhu2025revisiting,li2022confluence}.
These models are particularly relevant in settings involving nation-state actors, where cyber operations are embedded within broader geopolitical and economic considerations.

By framing agentic cyber conflict as a game between adaptive, learning players, one can reason not only about immediate best responses, but also about equilibrium behavior, incentive alignment, and stability under continual adaptation.
This perspective enables the design of defense strategies that are robust not just to current attacks, but to future adversarial learning and strategic evolution, which is an essential requirement for resilient cyber defense.

Beyond providing a lens for analyzing adversarial behavior, a game-theoretic framework also offers a principled pathway toward \emph{resilience}.
In adaptive attacker–defender settings, preventing every successful attack is neither realistic nor necessary.
Instead, the objective shifts to sustaining acceptable system performance over time despite persistent and intelligent adversarial pressure.
From this perspective, “winning” the cyber game means limiting attacker impact, preserving critical functionality, and enabling rapid recovery rather than achieving perfect prevention.

Game-theoretic reasoning naturally supports this shift by modeling repeated interaction, strategic adaptation, and incentive alignment \cite{zhu2012dynamic,zhu2020cross,zhu_basar_resilience2024}.
By explicitly accounting for how today’s defensive actions shape future attacker behavior, game-based approaches enable the design of strategies that stabilize long-term outcomes even as adversaries learn and evolve.
Resilience thus emerges not as a static property, but as a strategic equilibrium of ongoing cyber conflict, an outcome that is both achievable and essential in agentic security environments.

\begin{table}[t]
\centering
\caption{Agentic AI Across the Cyber Kill Chain}
\label{tab:killchain}
\renewcommand{\arraystretch}{1.25}
\begin{tabularx}{\textwidth}{p{3cm}X X}
\toprule
\textbf{Kill Chain Stage} &
\textbf{Agentic AI Role (Offense)} &
\textbf{Agentic AI Role (Defense)} \\
\midrule

Reconnaissance &
Autonomous scanning, system fingerprinting, and OSINT-driven intelligence
collection via semantic query generation and large-scale crawling. &
Continuous environment mapping, red-teaming simulation, and anticipatory
reconnaissance to minimize exposure and reduce attack surfaces. \\

Weaponization &
Automated exploit synthesis, malware generation, and payload optimization
leveraging generative models and reinforcement learning. &
Automated exploit reproduction, patch synthesis, and self-healing code
generation to neutralize discovered vulnerabilities. \\

Delivery &
Adaptive phishing and spearphishing campaigns with context-aware content
personalization that exploit cognitive and situational vulnerabilities. &
Intelligent content filtering, anomaly-based message vetting, and AI-driven
deception channels that disrupt or absorb delivery vectors. \\

Exploitation &
Real-time exploit chaining, privilege escalation, and lateral movement
coordinated by multi-agent policies and adaptive learning. &
Predictive isolation, sandboxing, and containment via behavioral modeling
and dynamic privilege reduction. \\

Installation \& Command-and-Control &
Autonomous deployment, persistence establishment, and distributed
command-and-control coordination. &
Rapid infection localization, recovery orchestration, and decentralized
counter-C2 mechanisms that sever attacker control. \\

Actions on Objectives &
Coordinated data exfiltration, system sabotage, or integrity and availability
manipulation across cyber--physical environments. &
Continuity assurance through redundancy, mission migration, and adaptive
failover enabled by predictive and resilient control. \\

Attacks Against Humans &
Cognitive manipulation using synthetic media, voice spoofing, and persuasive
dialogue systems targeting psychological biases. &
Cognitive security enhancement via AI-mediated training, adversarial
red-teaming, and human--AI trust calibration. \\

\bottomrule
\end{tabularx}
\end{table}

\section{Agentic AI for Cyber Resilience as the Next Frontier}

As agentic AI accelerates defense, it also accelerates the threat landscape.
Attackers can now leverage widely available AI tools to automate reconnaissance,
exploitation, and campaign orchestration at scales and speeds that were
previously impractical.
This paradigm shift requires a move beyond cybersecurity as prevention alone
toward \emph{cyber resilience}: the ability to absorb disruption, maintain
critical functions, recover rapidly, and learn from attacks.

Whereas security emphasizes \emph{prevention and protection}, resilience focuses
on \emph{continuity and recovery}.
It acknowledges that breaches are inevitable and that no system, regardless of
defensive sophistication, can remain impenetrable indefinitely.
A resilient system is therefore not defined by immunity to attack, but by its
capacity to sustain mission-critical functions, degrade gracefully, and recover
intelligently.
Failure is treated not as a terminal condition, but as a source of information
that drives adaptation and improvement.

The growing importance of resilience is driven by three converging realities.
First, the expansion of the attack surface through ubiquitous connectivity and
cyber--physical integration renders comprehensive protection infeasible.
Second, adversaries increasingly exploit human and cognitive vulnerabilities,
such as trust, bias, and attention, against which purely technical defenses offer
limited protection.
Third, autonomous and agentic AI systems capable of launching, coordinating, and
adapting attacks at machine speed fundamentally alter the tempo and structure of
cyber conflict.
Under these conditions, resilience becomes the defining attribute of sustainable
security: it enables systems to operate under uncertainty and persist through
continuous disruption.

Resilience does not replace security; rather, the two are complementary \cite{zhu_basar_resilience2024}.
Security mechanisms aim to prevent and detect intrusions, while resilience
mechanisms ensure recovery, adaptation, and continuity once defenses are
breached.
Because defensive resources, including computation, time, human attention, and operational
flexibility, are inherently finite, there exists a nontrivial tradeoff between
investment in prevention and investment in recovery.
Some vulnerabilities are best addressed through hardening, monitoring, and rapid
patching; others are more effectively managed through resilience strategies that
enable rapid reconfiguration or controlled degradation.
Determining this balance is not a matter of intuition, but of formal reasoning.

 \subsection{Facets of Cyber Resilience}
\label{subsec:facets_resilience}

Cyber resilience is inherently temporal.
It concerns not only whether a system can withstand a disruption at a given
instant, but how it behaves across the full lifecycle of adversarial interaction.
A resilient cyber system must therefore operate coherently before, during, and
after an incident, adapting its behavior as conditions evolve over time.

Before an incident occurs, the system must anticipate potential disruptions and
prepare to mitigate their impact.
During an incident, it must respond rapidly under uncertainty to preserve
mission-critical functionality.
After the incident, it must analyze what occurred, extract actionable insights,
and update its models, policies, and configurations to improve future
performance.
These temporal phases are distinct but tightly coupled: preparation shapes
response, response generates experience, and experience informs future
anticipation.

To capture this structure, cyber resilience can be decomposed into three
interdependent mechanisms: \emph{proactive}, \emph{responsive}, and
\emph{retrospective} resilience \cite{zhu2025foundations}.
Together, these mechanisms form the temporal backbone of adaptive cyber defense,
transforming resilience from a static property into a continuously evolving
capability.

\paragraph{Proactive resilience.}
Proactive resilience captures the system’s anticipatory capacity to foresee
potential disruptions and mitigate their impact before they occur.
It manifests at design and planning stages through architectural diversification,
redundancy, moving-target defenses, and strategic deception.
By embedding heterogeneity into components and pathways, the system reduces its
exposure to single points of failure and large-scale compromise.
Agentic AI augments proactive resilience by simulating adversarial behavior,
identifying high-risk attack surfaces, and autonomously deploying preemptive
countermeasures.
Through continuous monitoring and meta-learning, AI agents maintain a dynamic
understanding of threat evolution, enabling defenses to adapt even in the absence
of explicit attacks \cite{tao23meta, tao2024meta}.

\paragraph{Responsive resilience.}
Responsive resilience governs real-time adaptation once a disruption has occurred.
It encompasses rapid diagnosis, containment, and reconfiguration to maintain
essential operations under stress.
Traditional reactive defenses rely heavily on predefined playbooks and human
intervention; by contrast, agentic AI introduces adaptive, autonomous responses.
Intelligent agents can localize breaches, isolate compromised components, reroute
data flows, and reallocate computational resources in real time.
Operating under partial observability, these agents employ learning- and
control-based reasoning to balance containment speed against operational impact.
By distributing responses across multiple agents, the system achieves faster
reaction times and improved fault tolerance without centralized control.

\paragraph{Retrospective resilience.}
Retrospective resilience reflects the system’s capacity to learn and evolve after
an incident.
It includes forensic analysis, causal attribution, and post-event policy
refinement.
Agentic AI accelerates this process by reconstructing attack trajectories,
correlating indicators of compromise, and generating structured representations
(e.g., attack graphs or knowledge graphs) that encode causal dependencies.
From these representations, agents identify recurring weaknesses and update
defensive strategies accordingly.
This process transforms operational experience into structural knowledge,
closing the loop between failure and adaptation.

These three mechanisms form a continuous cycle of sensing,
reasoning, and acting across time.
Proactive anticipation informs real-time response, while the outcomes of the response
feed back into future anticipation and learning.
This recursive structure elevates resilience from a passive property to an active,
self-reinforcing process.
Agentic AI provides the cognitive substrate for this process by automating
perception, decision-making, and adaptation across temporal scales.
Through distributed intelligence and game-aware coordination, it enables cyber
systems to move beyond static defense toward dynamic resilience, which is an operating
regime in which security is not merely preserved, but continuously renewed
through interaction, learning, and strategic adaptation.

\subsection{Agentic AI Workflows for Cyber Resilience}

The notion of a \emph{security workflow} has traditionally referred to a sequence of operations that detect, analyze, and respond to cyber incidents. 
In classical architectures, such workflows are implemented as static, rule-based control loops, for example, the canonical \emph{detect--respond} cycle used in security operations centers (SOCs). 
While effective for routine events, these loops are fundamentally reactive and lack the capacity to generalize across dynamic and uncertain environments. 
They assume that threats evolve more slowly than the human decision cycle and that policy updates can be introduced manually. 
Such assumptions no longer hold in the era of autonomous, adaptive, and AI-generated attacks.

Agentic AI transforms these workflows from static pipelines into \emph{dynamic, adaptive ecosystems}. 
Instead of a single feedback loop between detection and response, agentic workflows comprise multiple interlocking loops of sensing, reasoning, and acting that operate concurrently across spatial and temporal scales. 
Each loop is managed by one or more AI agents that perceive their local environment, share contextual knowledge, and execute coordinated decisions. 
This distributed structure mirrors the principles of cybernetic control but extends them through machine learning and cooperative reasoning. 
The result is an architecture capable of continuous adaptation, self-healing, and anticipatory defense.

At the core of these workflows are three defining properties: autonomy, contextual awareness, and adaptive decision-making. 
\emph{High autonomy} refers to the ability of AI agents to function with minimal human supervision, executing end-to-end tasks that range from vulnerability discovery to remediation. 
Autonomy enables security systems to sustain operations under degraded conditions, respond faster than human analysts, and perform large-scale actions across distributed infrastructures. 
For example, an agentic patch management system can autonomously identify exploitable configurations, generate corrective code, and validate its impact using reinforcement signals derived from network performance metrics.

\emph{Contextual awareness} denotes the ability of agents to maintain continuity of understanding across tasks and over time. 
Traditional automation treats each task in isolation, relying on static signatures or short-term inputs. 
Agentic AI, by contrast, constructs persistent cognitive states that integrate historical evidence, environmental cues, and multi-agent interactions. 
Through memory architectures and knowledge representations such as semantic graphs, agents preserve and share context across detection, response, and recovery cycles. 
This continuity allows them to reason about causal dependencies; e.g., linking an anomalous network flow to a misconfiguration pattern observed hours earlier, and to coordinate responses that account for systemwide implications rather than localized anomalies.

\emph{Adaptive decision-making} encapsulates the ability to adjust strategies dynamically based on feedback and evolving objectives. 
Rather than relying on fixed response playbooks, agentic workflows employ learning algorithms that continuously refine their policies. 
Reinforcement learning, game-theoretic reasoning, and Bayesian inference allow agents to evaluate tradeoffs between competing goals such as containment speed, resource cost, and collateral impact \cite{kim-tao25quantize}. 
As a result, decisions evolve as the environment changes: defensive agents can escalate or de-escalate their actions depending on adversarial behavior, confidence levels, and mission priorities. 
This adaptivity transforms cybersecurity from a rule-based discipline into a dynamic control process driven by feedback and optimization.

Agentic workflows can be organized hierarchically to support resilience at multiple levels of abstraction. 
At the \emph{micro level}, operational agents handle immediate tasks such as packet inspection, anomaly scoring, and access verification. 
At the \emph{meso level}, coordination agents synthesize inputs from lower tiers, allocate resources, and harmonize local actions with global objectives. 
At the \emph{macro level}, strategic agents engage in meta-reasoning: they analyze long-term trends, evaluate performance metrics, and adjust the policies governing lower layers. 
This hierarchical composition produces a multi-scale control architecture analogous to those found in biological immune systems or distributed sensor networks, which are systems that achieve global robustness through the local intelligence of many autonomous entities.

These interwoven workflows embody the principles of \emph{resilient autonomy}. 
They are not pre-programmed routines but evolving processes capable of reorganizing themselves in response to internal degradation or external pressure. 
For instance, when a subset of agents is compromised, neighboring agents can infer the disruption, quarantine the affected region, and reallocate responsibilities through consensus protocols. 
Similarly, when environmental uncertainty increases, agents can switch from deterministic policies to probabilistic exploration, ensuring coverage across the threat space. 
In both cases, resilience emerges as an emergent property of adaptive coordination rather than as a static defensive layer.

\section{Toward a System Theory for Agentic AI Workflow Design}
\label{sec:system_theory_agentic}

Achieving the resilience-oriented goals outlined in the previous sections
requires more than isolated algorithmic advances.
It calls for a \emph{system theory} of agentic AI workflow design that clarifies
how perception, reasoning, memory, tools, and human interaction should be
structured, composed, and coordinated.
Without such principles, agentic AI systems risk becoming ad hoc collections of
components rather than coherent, robust decision-making architectures.

A central challenge is that agentic AI workflows can take fundamentally
different forms, depending on how intelligence, control, and feedback are
organized.
Before addressing questions of optimality, it is therefore necessary to
understand the \emph{structural archetypes} of agentic workflows and the design
tradeoffs they entail.
Figures~\ref{fig:workflow_simple}--\ref{fig:workflow_dynamic} illustrate four
canonical workflow patterns that arise in practice.

\subsection{Workflow Archetypes}

\begin{figure}[t]
  \centering
  \includegraphics[width=0.4\linewidth]{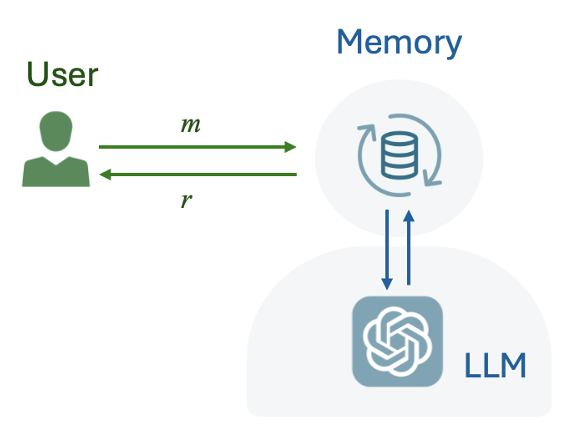}
  \caption{Simple agent-in-the-loop workflow, where an LLM-assisted agent mediates
  between a human user and external tools through prompt-driven interaction,
  without persistent memory or long-horizon adaptation.}
  \label{fig:workflow_simple}
\end{figure}

\paragraph{Simple agent-in-the-loop workflows.}
Figure~\ref{fig:workflow_simple} depicts the simplest form of an agentic workflow,
which closely resembles an enhanced chatbot interaction between a human user and
an LLM.
In this setting, the agent is primarily realized through \emph{prompt engineering}:
user instructions, contextual prompts, and tool descriptions are composed to guide
the model’s responses.
The LLM interprets user intent, optionally invokes external tools, and returns
generated outputs, but it does so within a single-turn or loosely coupled
multi-turn interaction pattern.

Functionally, the agent acts as a conversational interface that mediates between
the user and underlying computational resources.
There is no explicit separation between reasoning, memory, and control, and any
context required for decision-making must be reintroduced through prompts at each
interaction.
As a result, state persistence is shallow, implicit, and fragile, relying on
prompt context rather than structured memory or internal state evolution.

Such workflows are effective for information retrieval, exploratory analysis, and
lightweight decision support, where tasks can be completed within short
interaction horizons.
However, they lack persistent memory, long-horizon planning, and closed-loop
feedback with the environment.
Consequently, behavior remains largely reactive and myopic, with no principled
mechanism for learning from outcomes, adapting strategies over time, or sustaining
autonomous operation. 

\begin{figure}[t]
  \centering
  \includegraphics[width=0.6\linewidth]{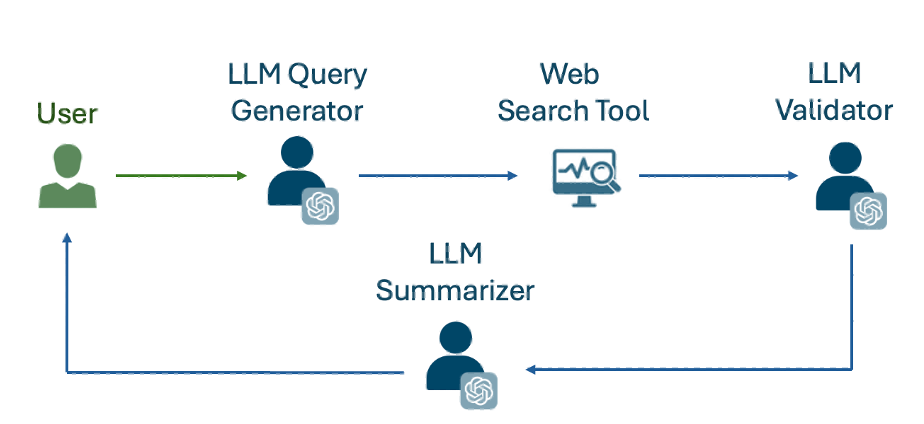}
  \caption{Static multi-stage agentic workflow with predefined roles and fixed
  control flow, suitable for structured tasks but limited in adaptivity under
  changing or adversarial conditions.}
  \label{fig:workflow_static}
\end{figure}

\paragraph{Static multi-stage workflows.}
Figure~\ref{fig:workflow_static} illustrates a more structured class of agentic
workflows, in which multiple agents or agentic components are composed into a
predefined sequence of stages.
Common examples include pipelines for query generation, external information
retrieval, validation, and summarization, where each stage performs a specialized
function and passes its output downstream.
This decomposition improves modularity, interpretability, and task-level
performance compared to single-agent workflows.

Despite these advantages, the defining characteristic of static multi-stage
workflows is that their control flow and role assignment are fixed at design time.
The workflow topology does not change in response to environmental feedback,
unexpected outcomes, or adversarial manipulation.
Agents execute prescribed roles without the ability to reallocate authority,
skip stages, revisit earlier decisions, or introduce new subtasks dynamically. 
If upstream assumptions are violated or downstream conditions change, the workflow
lacks mechanisms for self-correction beyond human intervention.
Consequently, static workflows remain effective for well-structured, predictable
tasks, but struggle in domains such as cybersecurity, resilience engineering, or
strategic planning, where uncertainty, deception, and adaptation are intrinsic.
From a system-theoretic viewpoint, these workflows represent a transitional stage
between scripted automation and fully dynamic agentic systems, offering structure
without true autonomy.

\begin{figure}[t]
  \centering
  \includegraphics[width=0.7\linewidth]{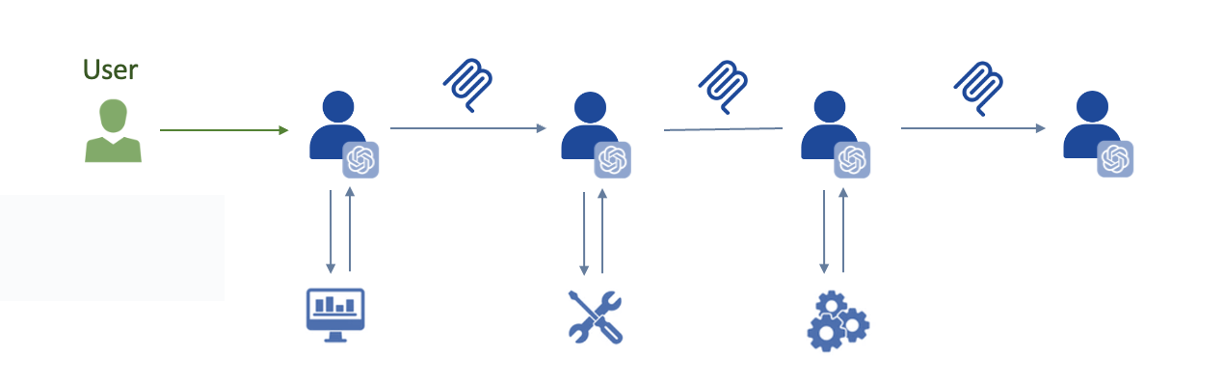}
  \caption{Decentralized sequential agentic workflow, where multiple agents perform
  localized reasoning and tool interaction in a staged, handoff-driven sequence.}
  \label{fig:workflow_dec}
\end{figure}

\paragraph{Decentralized sequential workflows.}
Figure~\ref{fig:workflow_dec} depicts a decentralized sequential agentic workflow,
in which multiple agents operate in a staged manner, each performing localized
reasoning and invoking tools relevant to its assigned subtask.
Information and intermediate results are passed sequentially from one agent to the
next, forming a pipeline of agentic decision-making rather than a centrally
orchestrated process. In this architecture, decision authority is distributed across agents, with each
agent responsible for interpreting its inputs, interacting with tools, and
producing outputs that condition subsequent stages.
This decomposition improves modularity, scalability, and fault isolation, as
individual agents can be modified or replaced without redesigning the entire
workflow.

\begin{figure}[t]
  \centering
  \includegraphics[width=0.6\linewidth]{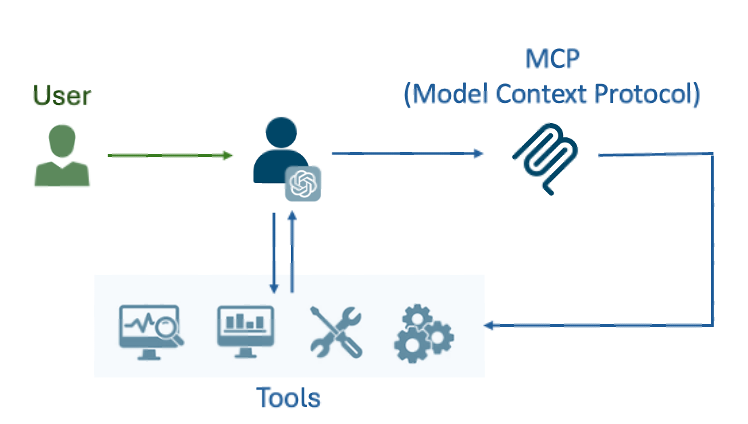}
  \caption{Dynamic closed-loop agentic workflow, in which agents interact with
  tools, memory, and the environment through continual feedback, enabling adaptive
  reasoning, action, and self-reconfiguration.}
  \label{fig:workflow_dynamic}
\end{figure}

\paragraph{Dynamic, closed-loop agentic workflows.}
Figure~\ref{fig:workflow_dynamic} illustrates fully dynamic agentic workflows in
which control flow is not pre-specified, but instead emerges from continuous
interaction among agents, tools, memory, and the environment.
Agents repeatedly sense system state, reason over evolving context, invoke tools,
and update internal memory, forming a closed-loop \emph{sense--reason--act--learn}
cycle.
Rather than progressing through a fixed sequence of stages, the workflow adapts
online, allowing agents to revise plans, switch roles, and reconfigure behavior
in response to feedback, uncertainty, and adversarial actions.

A key enabler of such workflows is the availability of a shared, structured
mechanism for managing and exchanging context across agents and tools.
Recent efforts, such as the \emph{Model Context Protocol (MCP)}, provide a unifying
interface for representing task state, intermediate results, tool affordances,
and environmental observations in a machine-interpretable form.
By externalizing and standardizing context, MCP-like protocols allow agents to
coordinate dynamically, invoke heterogeneous tools, and maintain coherent
long-horizon reasoning without relying on rigid orchestration logic.
This capability is essential for scaling agentic workflows beyond handcrafted
pipelines toward adaptive, self-directed systems.

Dynamic closed-loop workflows are particularly well-suited for adversarial and
resilience-critical settings.
In cybersecurity, they enable defenders to continuously adapt to attacker behavior
by revisiting hypotheses, re-prioritizing actions, and balancing competing
objectives such as containment, deception, and intelligence collection.
More broadly, these workflows support long-horizon planning, adaptive role
assignment, and autonomous recovery, making them a foundational architectural
pattern for agentic AI systems operating in nonstationary and contested
environments.

\subsection{Toward Optimal System-Level Design}
\label{subsec:optimal_system_design}

The value of a system-theoretic framework for agentic AI lies not only in its
ability to analyze adversarial interactions, but in its capacity to \emph{guide
the design of workflows themselves}.
In complex agentic systems, design choices, such as how many agents to deploy,
what roles they assume, how information and memory are distributed, and when
humans are engaged, fundamentally shape performance, robustness, and resilience.
Absent a principled framework, these choices are often made in an ad hoc manner,
hindering systematic comparison, optimization, and guarantees at the system
level.

\subsubsection{A Stackelberg Case Study for Optimal Agentic Workflow Design}
\label{subsec:stackelberg_penheal}

One illustrative case study of system-level agentic workflow design arises in
the context of \emph{automated penetration testing and remediation}, where multiple
agentic AI components must be coordinated to jointly optimize security and
resilience objectives \cite{huang2023penheal,huang4941478penhealnet}.
In contrast to monolithic automation, this setting naturally lends itself to a
\emph{game-theoretic formulation}, in which agents operate under asymmetric
information and sequential decision-making.
The \emph{PenHeal} framework provides a concrete example of how such interactions
can be structured and optimized using a Stackelberg game formulation.

PenHeal adopts a two-agent Stackelberg model in which decision-making unfolds
sequentially.
The first agent, the \emph{pentest agent}, explores the target system by emulating
attacker behavior and identifying exploitable vulnerabilities.
The second agent, the \emph{heal agent}, observes the pentest outcomes and
subsequently determines an optimal remediation strategy subject to operational
and cost constraints.
In this hierarchy, the heal agent assumes the role of the Stackelberg leader,
anticipating the information revealed by the pentest agent and selecting
countermeasures that maximize overall system security and resilience.
The pentest agent, in turn, functions as a follower whose exploration policy
reveals latent weaknesses in the system.

This sequential structure is critical from a workflow design perspective.
Rather than treating vulnerability discovery and remediation as loosely coupled
tasks, the Stackelberg formulation explicitly captures their \emph{strategic
interdependence}.
The exploration behavior of the pentest agent shapes the information available to
the heal agent, while the remediation actions chosen by the heal agent alter the
future attack surface and adversarial incentives.
By modeling this interaction explicitly, the workflow can be optimized at the
\emph{system level}, rather than by locally optimizing each agent in isolation.

Several important outcomes emerge from this two-agent design.
First, the coordinated Stackelberg workflow improves \emph{vulnerability
coverage}, as the pentest agent is incentivized to explore diverse attack paths
instead of repeatedly exploiting the same weaknesses.
Second, the heal agent’s optimization balances remediation effectiveness against
cost and operational disruption, selecting actions that provide maximal security
impact under resource constraints rather than defaulting to overly aggressive
patching.
Empirical results reported in the PenHeal study demonstrate that this strategic
coordination yields higher remediation effectiveness and reduced operational
overhead compared to non-strategic, decoupled baselines.

\subsubsection{Gestalt Game-Theoretic Framework for Agentic Workflow Composition}
\label{subsubsec:gestalt_framework}

The framework developed in our prior work~\cite{al2025gestalt} provides another
foundational approach for the \emph{system-level design of agentic AI workflows}
by modeling them as a \emph{coupled system of strategic decision processes}.
Rather than treating agents or tasks as isolated optimization problems, the
framework explicitly captures how local decision-making units are linked through
shared state evolution, information dependencies, and adversarial responses.
Each agent operates within a localized task environment under uncertainty and
strategic pressure, while its actions reshape the conditions faced by the downstream
agents and by the adversary itself.
This gestalt perspective enables reasoning about agentic workflows as coherent
\emph{systems of systems}, rather than as loosely connected pipelines of
automation.

From a workflow design standpoint, this formulation yields a set of actionable
system-level principles.
First, it provides a principled basis for determining \emph{where decision
authority should reside} within a multi-agent workflow.
Some decisions benefit from localized autonomy to enable fast reaction, scalability,
and robustness to partial failure, while others require coordination or
hierarchical oversight to maintain global consistency and strategic alignment.
The framework characterizes these tradeoffs by linking the strategic influence of
each decision point to its impact on overall system performance and adversarial
outcomes.

Second, the framework elevates \emph{information flow and memory placement} to
first-class design variables.
Rather than assuming uniform or unrestricted information sharing, it allows
designers to reason explicitly about what information should be retained locally,
aggregated across agents, or abstracted over time.
This distinction is particularly important for cyber resilience, where excessive
information propagation can enlarge attack surfaces or create single points of
failure, while insufficient sharing can degrade coordination, situational
awareness, and response quality.

Third, the framework supports the \emph{temporal composition of agentic functions}.
By modeling workflows as sequences of interdependent decision problems, it becomes
possible to analyze how proactive anticipation, real-time response, and
retrospective learning interact across time scales.
Designers can thus evaluate tradeoffs between early investment in sensing and
deception, rapid containment during incidents, and post-incident adaptation,
optimizing workflows for long-term resilience rather than short-term task success.

Crucially, optimality within this framework is defined at the \emph{system level}.
An individual agent’s policy is evaluated not solely by its local effectiveness,
but by how it contributes to the performance, stability, and resilience of the
workflow as a whole.
This naturally leads to equilibrium-based design criteria, in which workflows are
constructed such that no agent, or adversarial counterpart, can improve global
outcomes through unilateral deviation.
Such equilibrium notions provide robustness guarantees even in the presence of
adaptive and intelligent adversaries, aligning local decision-making with
long-horizon system objectives.

\subsubsection{Role of Game Theory in System-Level Agentic AI Design}
\label{subsubsec:game_theory_design}

Game theory plays a central role in system-level agentic AI design by providing a
\emph{unifying design language} that connects abstract workflow archetypes with
formal principles of strategic interaction.
As discussed earlier, agentic AI workflows can be organized into recurring
archetypes, such as static pipelines, sequential decision chains, dynamic
closed-loop workflows, and hierarchical or multi-agent compositions.
Each archetype embodies implicit assumptions about temporal ordering,
information availability, authority allocation, and adaptation \cite{tao_info}.
Game-theoretic models make these assumptions explicit and analytically tractable.

In particular, different workflow archetypes naturally map to different classes
of games.
Static or feedforward workflows correspond to one-shot or simultaneous-move
games, where agents act without observing downstream consequences.
Sequential workflows align with \emph{Stackelberg games}, in which leader–follower
structures encode temporal ordering and information asymmetry.
Dynamic, closed-loop workflows are naturally modeled as repeated or stochastic
games, capturing continual interaction, feedback, and learning over time.
Hierarchical and multi-agent workflows can be viewed as multi-level or nested
games, where local decision problems are embedded within higher-level strategic
objectives.

Within this framework, Stackelberg games are particularly important for agentic
AI workflow design.
They provide a principled mechanism for encoding who moves first, what
information is revealed, and how downstream agents and adversaries are expected
to respond.
This enables designers to explicitly reason about the placement of autonomy,
coordination, and human oversight within a workflow.
For example, early-stage agents may act as leaders who shape the information and
state space encountered by later agents, while downstream agents optimize
responses conditioned on those decisions.

Beyond specific game forms, equilibrium concepts provide optimality criteria at the system-level
 that transcend individual agents \cite{fudenberg1991game,bacsar1998dynamic}.
Rather than evaluating each agent individually, the game-theoretic design assesses
how agent policies jointly contribute to global objectives such as security,
resilience, and operational efficiency.
Equilibrium-based workflows are constructed so that no agent, or adaptive adversary, can improve overall outcomes through unilateral deviation, conferring
robustness under strategic uncertainty.

Crucially, this perspective elevates game theory from an analytical tool to an
\emph{architectural guide}.
It informs how agentic functions should be decomposed, how information and memory
should flow across workflow stages, and how the temporal and hierarchical structure
should be imposed.
By grounding agentic workflow archetypes in formal game-theoretic models,
designers can move from intuitive architectures to principled, optimizable, and
verifiable system-level designs \cite{zhu2025reasoning,zhu2025llm,zhu2025game,zhu2025generative,han2025dynamic}.

\section{Toward Agentic AI for Cyber-Physical Resilience}

The trajectory of cyber defense and resilience is increasingly shaped by the
convergence of \emph{agentic AI} and \emph{physical AI}.
This integration unites high-level computational cognition, including planning, reasoning,
and strategic adaptation, with embodied intelligence embedded in sensors,
actuators, robots, and control systems.
The result is a new class of \emph{agentic cyber--physical systems}, in which
intelligent agents operate simultaneously across digital and physical domains \cite{zhu2020cross,chen2022cross}.
In such systems, resilience is no longer confined to software or networks alone,
but becomes a property of the tightly coupled cyber--physical whole.

\begin{figure}[t]
  \centering
  \includegraphics[width=0.7\linewidth]{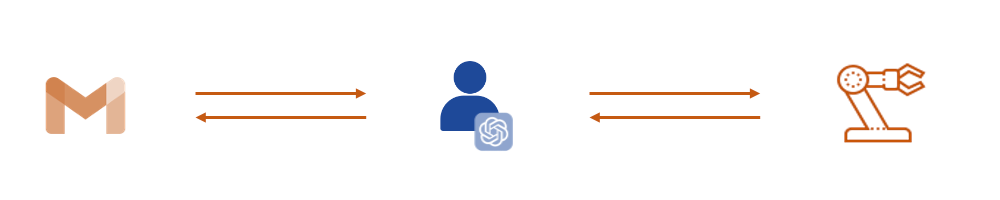}
  \caption{Convergence of agentic AI and physical AI in an integrated
  cyber--physical system. Agentic AI mediates bidirectional interaction between
  cyber services and physical systems, enabling closed-loop cognition, control,
  and adaptation.}
  \label{fig:agentic_cps}
\end{figure}

Figure~\ref{fig:agentic_cps} illustrates this convergence, where agentic AI
mediates bidirectional interaction between cyber services and physical systems.
On the cyber side, agentic AI interfaces with data platforms, communication
networks, and digital services.
On the physical side, it interacts with robots, autonomous vehicles, industrial
machines, and other embodied systems.
Through continuous feedback, the agentic layer interprets observations,
coordinates actions, and adapts control strategies, effectively becoming the
cognitive nexus where cyber intelligence meets physical execution.

The integration of cyber and physical AI is not incidental but necessary.
Modern critical systems, such as advanced manufacturing lines, autonomous aerial
vehicles, and cyber--physical infrastructure, are controlled by software-defined
logic while operating under physical constraints and safety requirements.
Cyber attacks in these environments can have direct physical consequences, while
physical disturbances or faults can propagate back into cyber layers.
Agentic AI provides the unifying decision-making substrate required to reason
across these domains, enabling coordinated responses that respect both digital
security objectives and physical safety constraints.

Consider a smart manufacturing system controlled by agentic and physical AI.
An adversarial cyber intrusion may attempt to manipulate sensor readings,
control commands, or production schedules \cite{manufacturing_5g_1,lee2015cyber,sahoo2022smart}.
An integrated agentic system can correlate cyber anomalies with physical signals
such as vibration patterns, energy consumption, or actuator behavior.
Rather than treating cyber and physical incidents separately, the agent reasons
about their joint impact, dynamically reconfiguring workflows, isolating
compromised components, and adjusting control policies to preserve production
quality and safety.
Resilience here emerges from the ability to maintain operational continuity even
as parts of the system are degraded or under attack.

A similar dynamic arises in autonomous UAV systems \cite{wang2025coordinated,zhang2017strategic,farooq2018multi,xu2015secure}.
Agentic AI oversees mission planning, coordination, and threat assessment, while
physical AI governs flight control, navigation, and collision avoidance.
Cyber attacks on communication links or onboard software can directly threaten
mission success and physical safety.
An integrated agentic cyber--physical architecture allows UAVs to adapt by
replanning routes, switching control modes, or transitioning to decentralized
coordination when connectivity is compromised.
In this setting, resilience is inseparable from the ability to reason jointly
about cyber threats, physical dynamics, and mission objectives.

At a systems level, resilience in integrated cyber--physical architectures arises
from the interaction of three tightly coupled loops.
The \emph{control loop} governs real-time physical stability, ensuring safe and
efficient operation of mechanical and electrical components.
The \emph{information loop} manages sensing, communication, and data integrity,
maintaining situational awareness across distributed components.
The \emph{cognitive loop}, orchestrated by agentic AI, reasons about intent,
anticipates adversarial strategies, and coordinates adaptive responses.
Although these loops operate at different time scales, their hierarchical
coupling enables the system to absorb disturbances, reconfigure under stress,
and recover functionality.

Such architectures can be naturally modeled using multi-layer game-theoretic and
hierarchical control frameworks \cite{zhu2025foundations}.
At lower layers, physical controllers stabilize dynamics against disturbances.
At higher layers, agentic decision-makers optimize strategies by anticipating
both adversarial behavior and environmental uncertainty.
Learning mechanisms continuously refine the coupling between perception, control,
and cognition, allowing resilience to emerge not from rigidity, but from
structured adaptivity.

The convergence of agentic AI and physical AI marks the point where
cyber resilience meets physical resilience.
It is at this interface that intelligent systems must defend not only data and
networks, but also safety, mission integrity, and physical continuity.
By embedding cognition into cyber--physical feedback loops, agentic AI enables a
form of resilience that is embodied, distributed, and adaptive, capable of
withstanding adversarial pressure in both digital and physical dimensions.

\section{Concluding Remarks and Future Prospects}
\label{sec:conclusion_future}

This chapter has argued that cybersecurity is entering a decisive transition
driven by foundation-model--based and agentic AI.
As autonomous reasoning, planning, and adaptation become accessible at scale,
both attackers and defenders increasingly operate as intelligent, learning
agents.
Under these conditions, security can no longer be treated as a static
configuration problem or a purely preventative exercise.
Instead, it must be understood as a dynamic, adversarial, and co-evolutionary
process.

We introduced the \emph{AI-augmented paradigm} as a conceptual lens for
understanding this shift and positioned \emph{agentic cyber resilience} as its
organizing principle.
Within this paradigm, resilience replaces absolute prevention as the primary
objective: systems are designed to anticipate disruption, maintain critical
functionality under attack, recover efficiently, and learn continuously.
Agentic AI enables this transformation by embedding sensing, reasoning, action,
and memory into closed-loop workflows that operate at machine speed while
remaining strategically aware.

A central contribution of the chapter is the development of a system-level view
of agentic AI workflow design.
Rather than optimizing isolated components, we showed how workflows must be
designed as coupled decision systems whose performance emerges from the
interaction of agents, information flows, temporal structure, and adversarial
responses.
Game-theoretic frameworks provide a unifying design language for this task,
allowing autonomy allocation, information disclosure, and temporal ordering to be
treated as explicit design variables.
Through case studies in cyber deception, penetration testing, and remediation, we
demonstrated how equilibrium-based reasoning leads to workflows that are not only
effective, but robust under strategic adaptation.

Looking forward, several open directions warrant further investigation.
First, the integration of agentic AI with large-scale cyber--physical systems
raises new challenges for resilience across digital, physical, and human layers.
Manufacturing systems, autonomous vehicles, energy infrastructure, and robotic
fleets increasingly rely on tightly coupled cyber and physical intelligence,
making coordinated resilience a first-class concern.
Second, the governance, verification, and assurance of autonomous agentic
workflows remain largely unexplored.
As decision authority shifts toward AI systems, new methods are required to
ensure accountability, safety, and alignment under adversarial pressure.

Finally, human–AI collaboration remains an essential dimension.
Designing workflows that balance autonomy with meaningful human oversight,
especially in high-stakes security and safety contexts, is a key challenge for
the next generation of resilient systems. Agentic AI should not be viewed merely as a more powerful automation
tool, but as a strategic actor embedded within complex socio-technical systems.
Cyber resilience in the era of AI-augmented reality will not be determined by eliminating
risk, but by how effectively systems can adapt, recover, and learn under
persistent adversarial pressure.
By grounding agentic workflow design in system theory and game-theoretic
principles, this chapter aims to provide a foundation for building security
systems that remain robust, adaptive, and resilient in the face of continual
change.

\bibliographystyle{abbrv}
\bibliography{ref}

\end{document}